\documentclass[aps,prl,twocolumn,floatfix,nofootinbib]{revtex4-1}
\usepackage{graphicx}
\usepackage{epic}
\usepackage{eepic}
\usepackage{latexsym}
\usepackage{amssymb,amsmath}
\usepackage{dutchcal}

\newcommand{\eq}[1]{(\ref{#1})}
\newcommand{\be}{\begin{equation}}
\newcommand{\ee}{\end{equation}}
\newcommand{\bea}{\begin{eqnarray}}
\newcommand{\eea}{\end{eqnarray}}

\newcommand{\hs}[1]{\hspace{#1 mm}}

\newcommand{\df}{\dot{\phi}}

\newcommand{\ca}{\mathcal{a}}
\newcommand{\cad}{\mathcal{\dot{a}}}

\def\a{\alpha}

\def\cc{\gamma}

\def\d{\delta}

\def\f{\phi}
\def\fr{\frac}
\def\F{\Phi}

\def\l{\lambda}
\def\L{\Lambda}

\def\s{\sigma}

\def\th{\theta}

\def\z{\zeta}

\def\del{\partial}

\let\bm=\bibitem
\def\nn{\nonumber}

\begin{document}

\title{Initial wave-function of the universe is arbitrary} 

\author{Ali Kaya}
\affiliation{Bo\~{g}azi\c{c}i University, Department of Physics, 34342, Bebek, \.{I}stanbul, Turkey}

\date{\today}

\begin{abstract}
	
We consider quantization of the gravity-scalar field system in the minisuperspace approximation. It turns out that in the gauge fixed deparametrized theory where the scale factor plays the role of time, the Hamiltonian can be uniquely defined without any ordering ambiguity as the square root of a self-adjoint operator. Moreover, the Hamiltonian degenerates to zero and the Schr\"{o}dinger equation becomes well behaved as the scale factor vanishes. Therefore, there is no technical or physical obstruction for the initial wave-function of the universe to be an arbitrary vector in the Hilbert space, which demonstrates the severeness of the initial condition problem in quantum cosmology. 

\end{abstract}

\maketitle

Although quantum mechanics has a lot of intriguing features, the evolution dictated by the Schr\"{o}dinger equation is fully deterministic, which is no more different than classical wave propagation. The wave-function is uniquely determined in time once the initial state is somehow specified at the beginning. In principle, the initial state can be arbitrarily chosen by the external agent preparing the system inasmuch as the freedom of choosing initial conditions in classical mechanics. 

The situation seems to be different in an approach to quantum cosmology based on the Wheeler-DeWitt (WDW) equation due to its timeless characteristic. Some would argue that timelessness is a fundamental property of quantum gravity. One then hopes to impose certain (mathematically and physically motivated) boundary conditions for the WDW equation in quantum cosmology that are supposed to fix the wave-function of the universe uniquely. This, however, is not easily achievable and there are different viable suggestions like the no boundary \cite{nb} or the tunneling \cite{t} proposals (see also \cite{ot1,ot2} for other alternatives and \cite{boj} for an attempt to determine the initial state by referring to the special dynamics of the loop quantum cosmology). 

More recently, a Lorentzian path integral method based on the Picard-Lefschetz theory has been put forward as the basis of quantum cosmology and it has been claimed that both the no boundary and the tunneling proposals yield unsuppressed perturbations, thus they are problematic in the Lorentzian framework \cite{mr1,mr2,mr3,mr4}. This has raised a debate in the literature, see \cite{s1,s2,s3,s33,s55,s4,s5,s6}. The wave-function obtained by the Lorentzian path integral solves the WDW equation, at least in the minisuperspace model when the proper-time gauge is employed \cite{hal}, and the recent discussion is again centered around the WDW approach.  

General relativity has many peculiarities that clearly distinguish it from a standard gauge theory of internal symmetries. To begin with the spacetime itself becomes dynamical rather than being a fixed background. Somehow relatedly, the ``structure constants" of the gauge algebra of coordinate transformations turn out to be field dependent. Moreover, the Hamiltonian of general relativity vanishes and it becomes a constraint of the theory. Nevertheless, all these properties do not imply a perfect timelessness and indeed general relativity admits an initial value formulation just like any other classical field theory. The absence of such a formulation would be a great drawback because, as the main observers of the universe, we definitely notice time evolution. 

One may possibly quantize a gauge theory by first eliminating all redundant degrees of freedom, i.e. by fixing a gauge and solving the constraints, which only leave a set of basic variables to quantize (this is sometimes called deparametrization). Naturally, physics should not change for different gauge choices that may yield distinct variables. Although it may be difficult to show this equivalence, it is just a matter of consistency which must be satisfied by any reasonable theory describing nature. Hence, this is a perfectly valid approach which can also be applied to quantize gravity. 

In this letter, we consider the system of a real self-interacting scalar field minimally coupled to gravity in the minisuperspace approximation, which is the standard setup for slow roll inflation. This is a well known toy model of quantum cosmology; while the {\it massless} deparametrized theory is worked out in \cite{sf0}, the WDW framework is studied in \cite{sf}. Here, we would like to consider the quantization of the deparametrized theory in a suitable gauge that generalizes the work of \cite{sf0} to a massive scalar field. As we will see, this leads to an ordinary one-dimensional quantum mechanical system with a perfectly defined Hamiltonian, which is free from any ordering ambiguities and gives well defined time evolution. This should be compared to the timeless WDW description that suffers from ordering issues. Surprisingly, the Schr\"{o}dinger equation turns out to be smoothly extendable through the big-bang singularity, so the model offers a way of resolving it in a simple manner. 

To see these interesting features, let us start from the following standard action of gravity-scalar field system
\be\label{s0}
S=\int d^4 x\sqrt{-g}\left[R-\fr12(\del\f)^2-V(\f)\right],
\ee
where we set the Planck mass $M_p^2\equiv1/(16\pi G)=1$. For the minisuperspace approximation, we take the metric as 
\be
ds^2=-N^2dt^2+a^2d\vec{x}^2\label{m1}
\ee
and assume that all fields $N$, $a$ and $\f$  depend only on time $t$. The Ricci scalar of \eq{m1} can be calculated as 
\be
R=-6\fr{\dot{a}}{a}\fr{\dot{N}}{N^3}+\fr{6}{N^2}\left(\fr{\dot{a}^2}{a^2}+\fr{\ddot{a}}{a}\right),
\ee
where the dot denotes the time derivative. Using these in \eq{s0} and applying an integration by parts give the standard minisuperspace action
\be\label{s1}
S=\int\, dt\, a^3\left[-\fr{6}{N}\fr{\dot{a}^2}{a^2}+\fr{1}{2N}\dot{\f}^2-NV\right].
\ee
The trivial integration over the spatial coordinates $\vec{x}$ in \eq{s0} yields the volume of space ${\cal V}$ as an overall term in the action, which can be absorbed in the scale factor so that 
\be
a\to ({\cal V})^{1/3}\, a.
\ee
Thus, the mass dimensions of the variables can be identified as  
\be
[N]=M^0,\hs{3}[a]=M^{-1},\hs{3}[\f]=M.
\ee
The action \eq{s1} has the following gauge invariance up to surface terms
\be\label{gtrans}
\d N=k^0\dot{N}+N\dot{k}^0,\hs{3}\d a=k^0\dot{a},\hs{3}\d \f=k^0\dot{\f},
\ee
where $k^0$ refers to a purely timelike diffeomorphism. The corresponding Noether current, which actually becomes the Hamiltonian in the canonical formulation, can be calculated as 
\be
j^0=-\fr{6}{N}a\dot{a}^2+\fr{1}{2N}a^3\df^2+a^3NV.
\ee
As discussed in \cite{ali}, the invariance of the action up to surface terms is enough to guarantee the existence and the conservation of the current, here $dj^0/dt=0$, and $j^0$ is the symmetry generator both in the classical and in the quantum theories. 

To fix the gauge symmetry we will impose the following gauge condition
\be\label{aa}
G\equiv a-\ca(t)=0,
\ee
where $a$ is the dynamical variable and $\ca(t)$ is an arbitrary (but fixed) positive function obeying $\cad(t)>0$ (see \cite{oa} for an alternative approach). In fact, this gauge breaks down when $\dot{a}=0$, but if one assumes $\dot{a}>0$ then for any $a$ one can apply the gauge symmetry \eq{gtrans} with a suitable $k^0$ to transform $a$ to the given $\ca(t)$, which shows that \eq{aa} is a proper gauge choice.\footnote{To see that \eq{aa} is an allowed gauge condition for $\dot{a}>0$ in an alternative way, one can also calculate its Poisson bracket with the constraint $\F$ given in \eq{nf} and check whether this vanishes or not. One may find that $\{G,\F\}=-P_a/(12a)\not=0$ provided $P_a\not=0$ or equivalently $\dot{a}\not=0$.}

To ensure that $\dot{a}>0$, we assume $V(\f)>0$ which divides the constrained phase space into two {\it disconnected subspaces} with $\dot{a}>0$ and $\dot{a}<0$. Eq. \eq{aa} is applicable in the subspace\footnote{If $V(\f)=0$ say at $\f=0$, then $\dot{a}=0$ defines a line given by $(a,P_a,\f,P_\f)=(a,0,0,0)$ in the 3-dimensional constrained phase space. Therefore one can still use \eq{aa}, for instance in the path integral, since the set of all configurations passing through that line is presumably of measure zero and thus negligible. Still, we are not going to study this case here.} with $\dot{a}>0$. We note that one usually employs the proper-time gauge $\dot{N}=0$ in this minisuperspace model, which is appropriate for the WDW formalism, see e.g. \cite{hal}. However, the proper-time gauge is not suitable for deparametrization since $N$ is already a non-propagating Lagrange multiplier. 

Let us now discuss how gauge fixing can be done in the Lagrangian and the Hamiltonian formalisms. In the Lagrangian approach, one can solve for $N$ from its own equation of motion as	 
\be\label{n}
N=\fr{1}{\sqrt{V(\f)}}\sqrt{6\fr{\dot{a}^2}{a^2}-\fr12 \df^2}.
\ee
Note that this solution demands the expression inside the square root to be non-negative which is consistent with the constrained phase space in which the dynamics occurs. Since the solution is algebraic, one may use it back in the action \eq{s1} that gives
\bea
\mathcal{S}&&=-2\int dt\, a^3\,\sqrt{V(\f)}\,\sqrt{6\fr{\dot{a}^2}{a^2}-\fr12 \df^2},\nn\\
&&=-2\int d\ca \ca^2\,\sqrt{V(\f)}\,\sqrt{6-\fr{\ca^2}{2} \left(\fr{d\f}{d\ca}\right)^2}\label{la}.
\eea
where in the second line we impose the gauge \eq{aa}. Here, the scale factor disappears as a dynamical variable and plays the role of time, and the scalar $\f(\ca)$ survives as the only degree of freedom. Note that the arbitrariness related to the choice of $\ca(t)$ in \eq{aa} also goes away automatically. 

For the gauge fixing in the Hamiltonian formalism, one can first determine the conjugate momenta from \eq{s1} as 
\be\label{pp}
P_a=-\fr{12}{N}a\dot{a},\hs{3}P_\f=\fr{1}{N}a^3\df.
\ee
We treat $N$ as a Lagrange multiplier and thus apply the Legendre transformation to other variables which gives the Hamiltonian as
\be\label{nf}
H=N\left[-\fr{1}{24a}P_a^2+\fr{1}{2a^3}P_\f^2+a^3V(\f)\right]\equiv N\Phi.
\ee
The variation of $N$ imposes the constraint $\Phi=0$ which can be solved to determine $P_a$ in terms of other variables, 
\be
P_a=\pm\fr{\sqrt{12}}{a}\sqrt{P_\f^2+2a^6V(\f)}.
\ee
Imposing $dG/dt=\{G,H\}+\del G/\del t=0$ fixes the lapse $N=-12a\cad/P_a$. As pointed out above, the constrained phase spaces becomes the disjoint union of two subspaces with $P_a>0$ and $P_a<0$ when $V(\f)>0$. Using the solution with $P_a<0$ that gives $\dot{a}>0$ and imposing the gauge \eq{aa} in \eq{s1} yield the following reduced action in the Hamiltonian form
\be\label{ha}
\mathcal{S}=\int d\ca\left[P_\f \fr{d\f}{d\ca}-\fr{\sqrt{12}}{\ca}\sqrt{P_\f^2+2\ca^6V(\f)}\right].
\ee
Again there remains a single degree of freedom described by the canonical variables $(\f,P_\f)$ and the scale factor plays the role of time. One can easily check that the gauge fixed actions \eq{la} and \eq{ha} are related by a Legendre transformation. 

It is instructive to see how the above gauge fixing can be carried out in the canonical path integral approach. Among other things this derivation clarifies how the range of $N$ integration must be chosen (this happens to be a source of debate in the literature). The unconstrained phase space path integral involves 
\be
\int Da\, DP_a\, D\f\,DP_\f\, DN\, e^{iS},
\ee
where the action must be expressed in terms of the canonical variables as 
\be
S=\int dt\left[P_a\dot{a}+P_\f\df-N\F\right].
\ee
The integration over $N$ should produce a Dirac delta functional $\d(\Phi)$ to confine the remaining integrals on the constrained phase space. In that case $N$ must certainly be integrated over $(-\infty,+\infty)$ even though in the classical theory it can be defined in the range $(0,+\infty)$. For gauge fixing one must include $\d(G)$ imposing the gauge condition and the accompanying Faddeev-Popov determinant $det\left\{G,\F\right\}$. In the subspace $P_a<0$ corresponding to $\dot{a}>0$, a simple calculation then gives 
\bea
\int Da\, DP_a\, D\f\,DP_\f\, \d(G)\,\d(\F)\,det\left\{G,\F\right\} e^{iS}\nn\\
=\int D\f\,DP_\f \,e^{i\mathcal{S}},
\eea
where $\mathcal{S}$ is given in \eq{ha}. 

As a result, we end up with a one-dimensional system governed by the Hamiltonian
\be\label{ch}
\mathcal{H}=\fr{\sqrt{12}}{\ca}\sqrt{P_\f^2+2\ca^6V(\f)},
\ee
which explicitly depends on ``time'' $\ca$. For quantization\footnote{The quantization of this system for $V(\f)=0$ was first studied in \cite{sf0}.} one can introduce the Hilbert space of square integrable functions with the standard inner product
\be
\left<\psi|\l\right>=\int_{-\infty}^{\infty}\,d\f\,\psi^*(\f)\,\l(\f)
\ee 
and represent the momentum operator as $P_\f=-i\del_\f$. If $V(\f)\to\infty$ as $\f\to\pm\infty$ (and $V(\f)>0$ as we have already assumed) then the essentially self-adjoint operator $P_\f^2+2\ca^6V(\f)$ will have a discrete {\it positive definite} point spectrum (for any given $\ca$) and therefore its square root giving the Hamiltonian in \eq{ch} becomes a well-defined operator which can easily be expressed in the corresponding orthonormal eigenbasis. The Schr\"{o}dinger equation 
\be\label{sce}
i\del_\ca\psi=\mathcal{H}\psi
\ee
can be formally solved by introducing the unitary evolution operator
\be\label{u}
U(\ca_2,\ca_1)=T\exp\left[-i\int_{\ca_1}^{\ca_2}\mathcal{H}(\ca)d\ca\right],
\ee
where $T$ denotes time ordering. All these are completely valid for any $\ca>0$.  

One may think that the Hamiltonian \eq{ch} becomes problematic at $\ca=0$, yet one should recall that it is implicitly defined as the square root of another operator whose spectrum happens to degenerate as $\ca\to0$. Thus, it might be possible to take $\ca\to0$ limit meaningfully by referring to a regular basis in the Hilbert space. To demonstrate this in an explicitly solvable example let us take 
\be\label{pot}
V(\f)=\fr12 m^2\f^2+\L,
\ee
where $\L>0$ is a constant that ensures $V(\f)>0$. In that case $\mathcal{H}$ becomes the square root of the harmonic oscillator Hamiltonian and the corresponding instantaneous orthonormal set of eigenfunctions are given by
\be\label{ef}
\psi_n=\fr{\sqrt{\a}}{\sqrt{2^nn!\sqrt{\pi}}}\,H_n(\a\f)\,e^{-\fr12\a^2\f^2},\hs{3}n=0,1,2...
\ee
where $\a=\sqrt{m\ca^3}$ and $H_n$ are Hermite polynomials. They satisfy $\left<\psi_n|\psi_m\right>=\d_{nm}$ and 
\be\label{hpsi}
\mathcal{H}\psi_n=E_n\psi_n,\hs{2}E_n=\fr{\sqrt{24}}{\ca}\sqrt{m\ca^3\left(n+\fr12\right)+\ca^6\L}.
\ee
Note that $\psi_n$ has explicit time dependence through $\a$. 

Let us first work out the evolution of states in the basis \eq{ef}. Expanding the wave-function as
\be
\psi=\sum_{n=0}^\infty c_ne^{-i\th_n}\psi_n,
\ee
where $d\th_n/d\ca=E_n$, the Schr\"{o}dinger equation \eq{sce} can be seen to imply
\bea
\fr{dc_m}{d\ca}=&&\fr{3}{4\ca}\left[\sqrt{m(m-1)}e^{i(\th_m-\th_{m-2})}c_{m-2}\right.\nn\\
&&\left.-\sqrt{(m+2)(m+1)}e^{i(\th_m-\th_{m+2})}c_{m+2}\right].\label{c}
\eea
One can check that the norm $\left<\psi|\psi\right>=\sum_n|c_n|^2$ is preserved, as it should be, but the phases of $c_n$ diverge logarithmically as $\ca\to0$. On the other hand, the basis \eq{ef} to which $c_n$ components refer are also ill defined at $\ca=0$. Hence, these equations can only be used for $\ca>0$.

To study the evolution in an entirely regular basis one can take \eq{ef} at a fixed nonzero $\ca$, say at $m\ca^3=1$, which gives
\be\label{el}
\l_n=\fr{1}{\sqrt{2^nn!\sqrt{\pi}}}\,H_n(\f)\,e^{-\fr12\f^2},\hs{3}n=0,1,2...
\ee
These also form an orthonormal set of basis vectors in the Hilbert space $\left<\l_n|\l_m\right>=\d_{nm}$, though they are not eigenstates of $\mathcal{H}$ except at $m\ca^3=1$. The matrix entries of $\mathcal{H}$ in this basis can be calculated as
\be\label{28}
\mathcal{H}_{nm}\equiv\left<\l_n|\mathcal{H}|\l_m\right>=\sum_{k=0}^\infty\left<\l_n|\psi_k\right>\left<\psi_k|\l_m\right>E_k.
\ee
Using \eq{ef}, \eq{el} and $E_k$ given in \eq{hpsi} one finds (after inserting proper $M_p$ factors) that 
\be
\lim_{\ca\to0}\mathcal{H}_{nm}=(mM_p^2)\,\ca^2\left[C_{nm}+{\it O}(\ca)\right],
\ee
where $C_{nm}$ are $\ca$ independent. Thus, $\mathcal{H}$ actually degenerates to the zero operator as $\ca\to0$. This somehow counterintuitive behavior arises  because both the spectrum and the eigenfunctions of $\mathcal{H}$ vanish at $\ca=0$, see \eq{ef} and \eq{hpsi}. The Schr\"{o}dinger equation can be extended through the moment $\ca=0$ without any problem, namely once the initial state $\psi(0)$ is specified at $\ca=0$ its time evolution can be determined as 
\be
\psi(\ca)=U(\ca,0)\psi(0),
\ee
where $U(\ca_2,\ca_1)$, which is given in \eq{u}, can be calculated without any issues at $\ca_1=0$. The initial energy can be considered to vanish for any $\psi(0)$ since the Hamiltonian degenerates to zero, which provides a concrete realization of the zero energy universe hypothesis \cite{tr}. 

Can there be a way of determining $\psi(0)$ in a natural way? In this minisuperspace model, the ``big-bang" singularity at $\ca=0$ is resolved and the evolution is well defined for any state in the Hilbert space. Since the Hamiltonian degenerates as $\ca\to0$, there is no ground state to be picked up as a plausible initial wave-function, which is sometimes considered as the most natural choice. Consequently, there seems to be no technical or physical reason for $\psi(0)$ not to be an arbitrary state in the Hilbert space. Let us note that in this model $\ca=0$ becomes the boundary of the spacetime and it is not possible the extend the geometry beyond it. Specifically, no bouncing configuration is allowed in the constrained phase space since $\dot{a}$ never vanishes. 

One may see that a solution of \eq{sce} approximately obeys the WDW equation (with a specific ordering) $\F\psi=0$ provided that \be\label{con}
\del_\ca\mathcal{H}\ll\mathcal{H}^2.
\ee
From \eq{28}, the matrix entries can be shown to have the following asymptotic behavior
\be
\lim_{\ca\to\infty}\mathcal{H}_{nm}\to\sqrt{\fr{a\L}{m}}, 
\ee
therefore \eq{con} is satisfied and $\psi$ becomes an approximate solution of the WDW equation at large times. One should however notice that while a WDW wave-function $\Psi(a,\f)$ gives a two-dimensional timeless probability distribution $|\Psi(a,\f)|^2dad\f$ in $(a,\f)$ space, the wave-function $\psi(\ca,\f)$ yields a probability distribution $|\psi(\ca,\f)|^2d\f$ for $\f$ at any given time $\ca$. In any case, $\psi(\ca,\f)$ is also a wave-function of the minisuperspace quantum gravity rather than being a state on a fixed background since its evolution involves sum over geometries in terms of the scale factor. 

It is possible to generalize the above results to the generic (spatially flat) minisuperspace model where the metric can be written as 
\be
ds^2=-N^2dt^2+h_{ij}(t)\,dx^idx^j.
\ee
For this case we directly carry out the Hamiltonian analysis and it is convenient to introduce the following variables 
\be
h_{ij}=e^\z\,\cc_{ij},
\ee
where $\det \cc_{ij}=1$ and $h\equiv\det h_{ij}=e^{3\z}$. Similarly, the conjugate momenta can be decomposed as  
\be
\pi^{ij}=\fr13 h^{ij}\,\pi+e^{-\z}\,\s^{ij},
\ee
where $\pi=\pi^{ij}h_{ij}$ and $\s^{ij}h_{ij}=0$. One can straightforwardly check that $(\z,\pi)$ and $(\cc_{ij},\s^{kl})$ are canonically conjugate pairs obeying the following Poisson brackets 
\bea
&&\{\z,\pi\}=1,\hs{3}\{\z,\s^{ij}\}=0,\hs{3}\{\cc_{ij},\pi\}=0,\nn\\
&&\{\cc_{ij},\s^{kl}\}=\fr12\left(\d^k_i\d^l_j+\d^l_i\d^k_j\right)-\fr13\cc_{ij}\cc^{kl},\\
&&\{\pi,\s^{ij}\}=0,\hs{3}\{\s^{ij},\s^{kl}\}=\fr13 \cc^{kl}\s^{ij}-\fr13\cc^{ij}\s^{kl},\nn
\eea
where we define $\cc^{kl}$ to be the inverse of $\cc_{kl}$. 

The action written in the Hamiltonian form can be found as 
\bea
S&&=\int dt\left[\pi^{ij}\,\dot{h}_{ij}+P_\f\,\df-N\F\right],\nn\\
&&=\int dt\left[\pi\,\dot{\z}+\s^{ij}\,\dot{\cc}_{ij}+P_\f\,\df-N\F\right],\label{2act}
\eea
where the constraint becomes   
\bea
\F\,&&=h^{-1/2}\left[\pi^{ij}\pi_{ij}-\fr12 \pi^2+\fr12P_\f^2+h\,V\right],\nn\\
&&=e^{-3\z/2}\left[-\fr16 \,\pi^2+\s^{ij}\s_{ij}+\fr12 \,P_\f^2+\,e^{3\z}\,V\right].\label{cons2}
\eea
Here the indices of $\s^{ij}$ are lowered by $\cc_{ij}$. As in the one-dimensional case, if $V>0$ the constrained phase space becomes the disjoint union of two subspaces with $\pi>0$ and $\pi<0$ (note that $\s^{ij}\s_{ij}\geq0$). This allows one to choose the conjugate variable $\z$ as the time coordinate. After solving $\pi$ from the constraint \eq{cons2} for the negative root and using $\z$ as time, one may obtain the deparametrized action 
\be
S=\int d\z\left[\s^{ij}\fr{d\cc_{ij}}{d\z}+P_\f\fr{d\f}{d\z}-\mathcal{H}\right],
\ee
where
\be\label{ham2}
\mathcal{H}=\sqrt{6\,\s^{ij}\s_{ij}+3P_\f^2+6\,e^{3\z}\,V(\f)}.
\ee
Note that the gravitational and the scalar field degrees of freedom decouple from each other. In these variables, the big-bang occurs at $\z=-\infty$ and the system has a well defined Hamiltonian in that limit.\footnote{To obtain the one-dimensional Hamiltonian \eq{ch} from \eq{2act} one should first utilize the following canonical transformation $\z=2\ln a$, $\pi=aP_a/2$ and then apply gauge fixing which uses $a$ as time.} Although quantization of the metric variables is nontrivial (and there is now an ordering ambiguity), the main conclusion does not change, i.e. one can still choose an arbitrary initial state at $\z=-\infty$. 

Despite being elementary, the present work signifies the severeness of the initial condition problem for the universe. There is no known and tested principle which restricts the initial state of a system completely or at least sufficiently. On the contrary, all our experience with physics taught us that the initial conditions are free by their nature. 

One would expect the quantum theory of gravity to resolve the big-bang singularity by replacing it either with a regular bounce or with a smooth initial beginning. The question is whether this requires new physics, or more importantly, if new physics involve a novel understanding of initial conditions. In loop quantum cosmology, one finds a smooth bounce \cite{ash} and in our toy minisuperspace model we see a smooth beginning with arbitrary initial state. These examples show that quantum evolution is in principle extendable through the big-bang singularity without extra physical assumptions. If this finally turns out to be the case for the true quantum theory of gravity, then the initial condition problem in cosmology becomes intractable. 

It is also possible that extra conditions might be needed to resolve the big-bang singularity. For example in \cite{ot1} DeWitt imposes the WDW wave-function to vanish on singular (3-dimensional) regions for the WDW equation to be well behaved (in the minisuperspace case this corresponds to $\Psi(0,\f)=0$). Similarly, the no boundary and the tunneling proposals can be viewed as specific boundary conditions for the WDW equation. However, none of these suggestions yield a unique initial state and it is difficult to judge their validity since mathematical consistency seems to be the only criterion. Besides, imposing a specific law, which is only valid at one time back in the history of the universe, is questionable from a philosophical point of view. In quantum gravity, any new physics that is imposed to resolve the big-bang singularity must also be applicable to black-holes or other possible singularities like big-crunches or big-rips. This also indicates that if a new law ever emerges, it is not going to be about restricting the initial state of the universe. 

The potential \eq{pot} is suitable for slow-roll chaotic inflation, which needs the inflaton scalar to acquire large values $|\f|>M_p$ in the semi-classical regime. One usually imagines that the scalar field randomly fluctuates and the regions in which this condition is met inflate. Such a picture seems to avoid any naturalness problem related to the likelihood of initial conditions yielding inflation. Nevertheless, it is hard to quantify these arguments and the exactly solvable toy model studied in this letter offers a good opportunity to illustrate the difficulties. For instance, given a normalized state $\psi(\f)$ at some time, one may tend to identify the probability of inflation as $p=1-\int_{-M_p}^{M_p} |\psi(\f)|^2d\f$. One may then think that in a universe with many regions, inflation seems inevitable irrespective of the state chosen even if it gives $p\ll1$. However, inflation requires the inflaton to be semi-classically localized around a definite value which can then be used in the classical Einstein's equations. Therefore, it is not clear, given an arbitrary state, how the scalar field can be thought to acquire a value unless a measurement is done. To avoid this problem one may assume that a well peaked Gaussian wave-function centered around some $|\f|>M_p$ is a suitable state for inflation. In that case the probability of inflation would become a calculable quantity if the initial state had been identified, which unfortunately appears impossible as we have shown.

\end{document}